\documentstyle[12pt,dina4]{article}

\catcode`\@=11
\long\def\@makefntext#1{
\protect\noindent \hbox to 3.2pt {\hskip-.9pt
$^{{\ninerm\@thefnmark}}$\hfil}#1\hfill}		

\def\@makefnmark{\hbox to 0pt{$^{\@thefnmark}$\hss}}  

\def\ps@myheadings{\let\@mkboth\@gobbletwo
\def\@oddhead{\hbox{}
\rightmark\hfil\ninerm\thepage}
\def\@oddfoot{}\def\@evenhead{\ninerm\thepage\hfil
\leftmark\hbox{}}\def\@evenfoot{}
\def\sectionmark##1{}\def\subsectionmark##1{}}

\newcounter{sectionc}\newcounter{subsectionc}\newcounter{subsubsectionc}
\renewcommand{\section}[1] {\vspace*{0.6cm}\addtocounter{sectionc}{1}
\setcounter{subsectionc}{0}\setcounter{subsubsectionc}{0}\noindent
	{\normalsize\bf\thesectionc. #1}\par\vspace*{0.4cm}}
\renewcommand{\subsection}[1] {\vspace*{0.6cm}\addtocounter{subsectionc}{1}
	\setcounter{subsubsectionc}{0}\noindent
	{\normalsize\it\thesectionc.\thesubsectionc. #1}\par\vspace*{0.4cm}}
\renewcommand{\subsubsection}[1]
{\vspace*{0.6cm}\addtocounter{subsubsectionc}{1}
	\noindent {\normalsize\rm\thesectionc.\thesubsectionc.\thesubsubsectionc.
	#1}\par\vspace*{0.4cm}}
\newcommand{\nonumsection}[1] {\vspace*{0.6cm}\noindent{\normalsize\bf #1}
	\par\vspace*{0.4cm}}

\newcounter{appendixc}
\newcounter{subappendixc}[appendixc]
\newcounter{subsubappendixc}[subappendixc]

\renewcommand{\appendix}[1] {\vspace*{0.6cm}
        \refstepcounter{appendixc}
        \setcounter{figure}{0}
        \setcounter{table}{0}
        \setcounter{equation}{0}
        \renewcommand{\thefigure}{\Alph{appendixc}.\arabic{figure}}
        \renewcommand{\thetable}{\Alph{appendixc}.\arabic{table}}
        \renewcommand{\theappendixc}{\Alph{appendixc}}
        \renewcommand{\theequation}{\Alph{appendixc}.\arabic{equation}}
        \noindent{\bf Appendix \theappendixc #1}\par\vspace*{0.4cm}}

\def\abstracts#1{{

\centering{\begin{minipage}{12.2truecm}\footnotesize\baselineskip=12pt\noindent
	\centerline{\footnotesize ABSTRACT}\vspace*{0.3cm}
	\parindent=0pt #1
	\end{minipage}}\par}}

\renewenvironment{thebibliography}[1]
	{\begin{list}{\arabic{enumi}.}
	{\usecounter{enumi}\setlength{\parsep}{0pt}
\vspace{-5mm}
\setlength{\leftmargin 0.52cm}{\rightmargin 0pt}
	 \setlength{\itemsep}{0pt} \settowidth
	{\labelwidth}{#1.}\sloppy}}{\end{list}}

\newcounter{itemlistc}
\newcounter{romanlistc}
\newcounter{alphlistc}
\newcounter{arabiclistc}

\newcommand{\fcaption}[1]{
        \refstepcounter{figure}
        \setbox\@tempboxa = \hbox{\footnotesize Fig.~\thefigure. #1}
        \ifdim \wd\@tempboxa > 6in
           {\begin{center}
        \parbox{6in}{\footnotesize\baselineskip=12pt Fig.~\thefigure. #1}
            \end{center}}
        \else
             {\begin{center}
             {\footnotesize Fig.~\thefigure. #1}
              \end{center}}
        \fi}

\newcommand{\tcaption}[1]{
        \refstepcounter{table}
        \setbox\@tempboxa = \hbox{\footnotesize Table~\thetable. #1}
        \ifdim \wd\@tempboxa > 6in
           {\begin{center}
        \parbox{6in}{\footnotesize\baselineskip=12pt Table~\thetable. #1}
            \end{center}}
        \else
             {\begin{center}
             {\footnotesize Table~\thetable. #1}
              \end{center}}
        \fi}

\def\@citex[#1]#2{\if@filesw\immediate\write\@auxout
	{\string\citation{#2}}\fi
\def\@citea{}\@cite{\@for\@citeb:=#2\do
	{\@citea\def\@citea{,}\@ifundefined
	{b@\@citeb}{{\bf ?}\@warning
	{Citation `\@citeb' on page \thepage \space undefined}}
	{\csname b@\@citeb\endcsname}}}{#1}}

\newif\if@cghi
\def\cite{\@cghitrue\@ifnextchar [{\@tempswatrue
	\@citex}{\@tempswafalse\@citex[]}}
\def\citelow{\@cghifalse\@ifnextchar [{\@tempswatrue
	\@citex}{\@tempswafalse\@citex[]}}
\def\@cite#1#2{{$\null^{#1}$\if@tempswa\typeout
	{IJCGA warning: optional citation argument
	ignored: `#2'} \fi}}

 1
 1
 1

\font\ninerm=cmr9

\begin{document}


\vspace*{-10mm}
\hfill {\footnotesize KL--TH--95/16} \\[5mm]
\hfill {\footnotesize hep-th/9510046} \\[5mm]

\centerline{\normalsize\bf BRST--INVARIANT}
\centerline{\normalsize\bf APPROACH TO QUANTUM MECHANICAL
TUNNELING}
\baselineskip=16pt

\vspace{0.6cm}
\centerline{\footnotesize JIAN-GE ZHOU, F. ZIMMERSCHIED,
J.-Q.LIANG\footnote{On leave
from  Institute of Theoretical Physics,  Shanxi University, Taiyuan, Shanxi
030006,
P.\ R.\ China, and Institute of Physics, Academia Sinica, Beijing 100080, P.\
R.\ China}
and H.J.W. M\"ULLER-KIRSTEN}
\baselineskip=13pt
\centerline{\footnotesize\it Department of Physics,
University of Kaiserslautern, P.O.Box 3049}
\baselineskip=12pt
\centerline{\footnotesize\it 67653 Kaiserslautern, Germany}
\centerline{\footnotesize E-mail: zimmers@.physik.uni-kl.de}

\vspace*{1cm}
\abstracts{
A new approach with BRST invariance is suggested to cure the degeneracy problem
of ill defined path integrals in the path--integral calculation
of quantum mechanical tunneling effects  in which the problem arises due to
the occurrence of zero modes.
The Faddeev--Popov procedure is
avoided and the integral over the zero mode is transformed in a systematic way
into a well defined
integral over instanton positions. No special procedure has
to be adopted as in the Faddeev--Popov method in calculating the Jacobian of
the transformation. The quantum mechanical tunneling for the Sine--Gordon
potential is used as a test of the method and the width of the lowest
energy band is obtained in exact agreement with that of WKB calculations.}

\vspace{5mm}
\section{Introduction}

Quantum tunneling has attracted considerable interest because of its wide
application in areas ranging from condensed matter to high energy physics.
The instanton method is a powerful tool for the calculation of tunneling
effects. Recent interests in tunneling effects were initiated by the work
of Ringwald\cite{1}, who argued that although the cross section of the standard
electroweak theory is proportional to an exponentially small WKB suppression
factor, it is nevertheless rapidly growing with energy due to multiple
production of Higgs and vector bosons. However, it has been observed that
instantons interpolate between neighbouring vacua and satisfy vacuum
boundary conditions and therefore may not be  adequate for a description
of tunneling at high energy\cite{2}. Motivated by the instanton--induced
baryon--number violating processes the instanton method of quantum
tunneling has recently been extended to tunneling at finite, nonzero
energy by means
of nonvacuum or periodic instantons\cite{3} which have become the subject
of extensive investigation under the name of sphalerons\cite{4,5}.

It is well known that due to the translational invariance of the action,
functional
integrals in nonlinear field theory are not well defined when expanded about
the
classical solutions of the field equations. The translational symmetry results
in
zero eigenmodes of the second variation operator of the action which are a
reflection of the arbitrariness of locations in space of solitons or instantons
(in Euclidean time). Obviously physical quantities are independent of such
center--of--mass--like locations. A consequence of the existence of these
normalizable ``zero modes''
which belong to the spectrum of linearised fluctuations in the soliton's
background
is a divergence one encounters when quantizing
the theory in the background of these solitons or instantons. In essence the
change of variables in field theory converts the zero modes into corresponding
collective coordinates. However, calculation of the Jacobian of the
transformation of variables in the path integral formalism usually involves a
Faddeev--Popov procedure\cite{8,9}.

Here we report an alternative method to carry out the transformation of
variables and the evaluation of the related Jacobian in a systematic way with
BRST invariance, and we apply it to quantum tunneling processes.
The crucial point of the method is the exploitation of a shift invariance
of the fluctuation action.
As a test case
the quantum tunneling effect in $1+0$ dimensions is calculated for the
Sine--Gordon (SG) potential. In the usual collective coordinate method
\cite{10,11,12} time--dependent collective coordinates are associated with the
$D$ (spatial) dimensional (static) solitons indicating the motion of the kink
or
center of mass and one has to deal with phase--space functional integrals
in $1+D$ dimensions. Unlike this procedure our BRST invariant treatment
does not evoke the time--dependence in one higher dimension and works
on configuration--space path--integrals. Tunneling in the case of the
Sine--Gordon potential is itself an interesting subject in view of similar
features of the $O(3)$ nonlinear sigma model in the discussion towards
understanding instanton induced baryon--number violating processes
\cite{13,14}.

\section{Energy band structure for the SG potential and the transition
amplitude for quantum tunneling}

The $1+0$ dimensional Lagrangian we consider is
\begin{equation}
{\cal L} = \frac12 \left(\frac{d\phi}{dt}\right)^2 - V(\phi)
\label{2.1}
\end{equation}
with the SG potential
\begin{equation}
V(\phi) = \frac{1}{g^2} [1 + \cos(g\phi)]
\label{2.2}
\end{equation}
where $g>0$ denotes a dimensionless coupling constant. The classical
solution which extremizes the action is seen to satisfy the equation of
motion which after one integration with integration constant zero is
\begin{equation}
\frac12 \left(\frac{d\phi_c}{d\tau}\right)^2 + V(\phi_c) = 0
\label{2.3}
\end{equation}
with euclidean time $\tau = it$. Mass $m=1$ and natural units $c=\hbar=1$
are used throughout. The relevant classical solution which interpolates between
neighbourung vacua, say, $\phi_+=\frac{\pi}{g}$ and $\phi_-=-\frac{\pi}{g}$, is
\begin{equation}
\phi_c=\frac{2}{g}\sin^{-1}[\tanh(\tau + a)]
\label{2.4}
\end{equation}
where the integration constant $a$ is interpreted as the position of the
instanton.

In the following we consider the case of  very
high potential barriers with correspondingly small tunneling contributions to
the eigenvalues. Thus we suppose $|0\rangle_+$, $|0\rangle_-$ are
degenerate eigenstates in neighbouring wells with the same
energy eigenvalue ${\cal E}_0$ such that $\hat{H}^0 |0\rangle_\pm =
{\cal E}_0 |0\rangle_\pm$ where $\hat{H}^0$ is the Hamiltonian of the
harmonic oscillator as the zeroth order approximation of the system. The
degeneracy will be removed by the small tunneling effect which leads to the
level splitting. The eigenstates of the Hamiltonian $\hat{H}$ then become
\begin{equation}
|0\rangle_o = \frac{1}{\sqrt{2}}\left(|0\rangle_+-|0\rangle_-\right), \quad
|0\rangle_e = \frac{1}{\sqrt{2}}\left(|0\rangle_++|0\rangle_-\right), \quad
\label{2.9}
\end{equation}
with eigenvalues ${\cal E}_0 + \Delta{\cal E}$ and
${\cal E}_0 - \Delta{\cal E}$ respectively. $ \Delta{\cal E}$ denotes the
shift of one oscillator level. It is obvious that
\begin{equation}
_+\langle 0| \hat{H}-\hat{H}^0|0\rangle_- = \Delta{\cal E}.
 \label{2.10}
\end{equation}
These shifts embrace the energy bands which result from the
translational invariance of $V$.
We now calculate this energy resulting from tunneling with the instanton
method. The amplitude for a transition between neighbouring vacua due
to instanton tunneling can be written
\begin{equation}
\langle \phi_f = \frac{n\pi}{g},T | \phi_i = \frac{(n-1)\pi}{g},-T\rangle
\equiv K(\phi_f,T;\phi_i,-T) = \int_{\phi_i}^{\phi_f}{\cal D}\{\phi\}e^{-S}
\label{2.11}
\end{equation}
where
\begin{equation}
S = \int_{-T}^T \left[\frac12\left(\frac{d\phi}{d\tau}\right)^2
+V(\phi)\right] d\tau
\label{2.22}
\end{equation}
is the Euclidean action. In the large time limit $T\rightarrow\infty$, we have
\begin{eqnarray}
K & = & \sum_{n,n'} \langle \phi_f | n \rangle\langle n | e^{-2\hat{H}T} | n'
\rangle
\langle n' | \phi_i \rangle \nonumber \\
& \simeq &
\Psi_0(\phi_f)\Psi_0(\phi_i)e^{-2T{\cal E}_0}\sinh(2\Delta {\cal E} T).
\label{2.23}
\end{eqnarray}

\section{Translation invariance of action and zero modes}

As mentioned in the introduction the degeneracy of an action which possesses
a translation symmetry leads to ill defined functional integrals in
perturbation
expansion about the classical configuration $\phi_c$ since the symmetry
results in zero modes of the second variation operator of the action. We
consider this now in more detail. We expand $\phi(\tau)$ about the classical
trajectory $\phi_c(\tau)$ and so set
\begin{equation}
\phi(\tau)=\phi_c(\tau) + \chi(\tau)
\label{3.1}
\end{equation}
with the boundary conditions $\chi(T)=\chi(-T)=0$ for the fluctuation.
Substituting $\phi(\tau)$ of eq. (\ref{3.1}) into eq. (\ref{2.11}) and
retaining
only terms up to the second order in $\chi$ for the one--loop approximation
we obtain
\begin{equation}
K = e^{-S_c}I
\label{3.2}
\end{equation}
where the classical action is evaluated along the trajectory $\phi_c$ so that
\begin{equation}
S_c = \int_{-\infty}^\infty \left[\frac12\left( \frac{d\phi_c}{d\tau}\right)^2
+V(\phi_c)\right] d\tau = \frac{8}{g^2}.
\label{3.3}
\end{equation}
The fluctuation integral $I$ is seen to be
\begin{equation}
I = \int_{\chi(-T)=0}^{\chi(T)=0} {\cal D}\{\chi\} e^{-\Delta S}
\label{3.4}
\end{equation}
with the fluctuation action
\begin{equation}
\Delta S = \int_{-T}^T \chi\hat{M}\chi d\tau
\label{3.5}
\end{equation}
where
\begin{equation}
\hat{M} = -\frac12 \frac{d^2}{d\tau^2} + \frac12 \left( 1-
\frac{2}{\cosh^2(\tau-a)}\right)
\label{3.6}
\end{equation}
is the self--adjoint
operator of the second variation about the classical trajectory.
Expanding $\chi(\tau)$ in terms of normalized eigenfunctions of $\hat{M}$
we set
\begin{equation}
\chi(\tau) = \sum_m c_m \Psi_m
\label{3.7}
\end{equation}
where
\begin{equation}
\hat{M}\Psi_m = E_m\Psi_m.
\label{3.8}
\end{equation}
Changing the integration variables of (\ref{3.4})
to $\{c_m\}$, the functional integral $I$
can be formally evaluated to be
\begin{equation}
I = \left|\frac{\partial\chi(\tau)}{\partial c_m}\right| \prod_m
\left[ \frac{\pi}{E_m}\right]^{\frac12}
\label{3.9}
\end{equation}
which is seen to be divergent in view of the vanishing eigenvalue of the
zero mode, $E_0 = 0$. Since the transformation (\ref{3.7}) is linear the
Jacobian  $\left|\frac{\partial\chi(\tau)}{\partial c_m}\right|$ is constant
and
has therefore been factored out. To cure the problem one normally resorts
to the so--called Faddeev--Popov procedure\cite{7,9} in order to
transform the integral over the zero mode $c_0$ into the continuous integration
of a collective coordinate, which is the  instanton position $a$ in our
case.

\section{BRST invariance and ``gauge fixing''}

In the usual collective coordinate method, the essential ingredient is a
change of variables which is such that every collective coordinate which
is time--dependent is associated with a zero mode\cite{11,12}. In our
$(1+0)$ dimensional case the procedure is not appropriate. We therefore
adopt an alternative method\cite{14,15} which employs a BRST invariance in
dealing with the transformation of variables.  If we identify euclidean time
$\tau$ with a spatial coordinate, the equivalence of the present method with
that of a collective coordinate method with BRST invariance is similar to
that demonstrated in a previous paper\cite{16}.

After expansion about the classical trajectory $\phi_c$, the fluctuation action
still retains a shift symmetry expressed by the invariance
\begin{equation}
\Delta S(\chi') = \Delta S(\chi)
\label{4.1}
\end{equation}
where
\begin{equation}
\chi' = \chi + \frac{\partial \phi_c}{\partial a}
= \chi + \frac{\partial \phi_c}{\partial \tau}.
\label{4.2}
\end{equation}
In other words the action is invariant under a kind of ``gauge transformation''
of the fluctuation variable $\chi$. This is an important observation.
The key point of the BRST procedure is to
enlarge the number of degrees of freedom and invent a nilpotent symmetry
which mimics the structure of the gauge symmetry. Then one can achieve the
effects of gauge fixing without breaking the BRST invariance of the
system.  To achieve this we first enlarge the configuration space by
considering the parameter $a$ as a variable. We replace the transformation
(\ref{4.2}) by the introduction of new anticommuting variables $c$ and
$\bar{c}$
and a Nakanishi--Lautrup auxiliary variable $b$
(the latter in such a way that it implements the gauge fixing condition as its
equation of motion, i.\ e.\ $b=\int\chi\frac{d\phi_c}{da}d\tau$ in analogy with
the
implementation of the Lorentz gauge in QED)
together
with conjugate momenta
$P_a$, $\Pi_c$, $\Pi_{\bar{c}}$ such that
\begin{eqnarray}
\delta \chi & = & -c \frac{\partial \phi_c}{\partial a} \nonumber \\
\delta a & = & -c \nonumber \\
\delta \bar{c} & = & 2\pi b \nonumber \\
\delta b & = & 0 \nonumber \\
\delta c & = & 0.
\label{4.3}
\end{eqnarray}
The variables $a$, $c$, $\bar{c}$, $\chi$ and their conjugate momenta are
assumed to satisfy the canonical Poisson relations
\begin{eqnarray}
\{a, P_a \} & = &  1  \nonumber \\
\{\bar{c}, \Pi_{\bar{c}} \}_+ & = & 1 \nonumber \\
\{c, \Pi_c \}_+ & = & 1    \nonumber \\
\{\chi, \pi_{\chi} \}  & = & 1.
\label{4.4}
\end{eqnarray}
Then one finds that the BRST transformations can be generated by the
following BRST charge
\begin{equation}
\Omega = -cP_a - c\pi\frac{\partial \phi_c}{\partial a}
+ \Pi_{\bar{c}}b
\label{4.5}
\end{equation}
which is nilpotent. The following BRST invariant term may now be added to
the fluctuation Lagrangian to break the shift symmetry:
\begin{eqnarray}
L_B & = & \int \delta \left[ \bar{c} \frac{\partial \phi_c}{\partial a}
\chi \right] d\tau
- \pi b^2 \nonumber \\
& = & 2\pi b\int\frac{\partial \phi_c}{\partial a}\chi d\tau -
c\bar{c}\int \left[ \frac{\partial^2 \phi_c}{\partial a^2}\chi + \left(
\frac{\partial \phi_c}{\partial a}\right)^2\right]d\tau - \pi b^2
\label{4.6}
\end{eqnarray}
Eliminating the Nakanishi--Lautrup auxiliary variable by using its equation
of motion, we obtain the final expression of the fluctuation functional
integral
\begin{eqnarray}
I & = & \int {\cal D}\{\chi\} {\cal D}\{c\} {\cal D}\{\bar{c}\} {\cal D}\{a\}
\nonumber \\
& & \cdot  \exp\left\{-\int\chi\hat{M}\chi \;d\tau -
\pi \left[ \int\frac{\partial \phi_c}{\partial a}\chi d\tau \right]^2 -
c\bar{c}\int\left[\frac{\partial^2 \phi_c}{\partial a^2}\chi + \left(
\frac{\partial \phi_c}{\partial a}\right)^2\right]d\tau \right\}.
\label{4.7}
\end{eqnarray}
(In the following we neglect the term linear in the fluctuation $\chi$
since it is of higher order than the soliton mass $M=\int d\tau \left(
\frac{d\phi_c}{da}\right)^2$).
Expanding the fluctuation variable $\chi$ in terms of the eigenmodes of
$\hat{M}$ so that
\begin{equation}
\chi = \sum_m c_m\Psi_m, \quad {\cal D}\{\chi\} =
\left|\frac{\partial\chi}{\partial c_m}\right|{\cal D}\{c_m\}
\end{equation}
we obtain (ignoring the higher order
contribution proportional to $\chi$ in the coefficient
of $c\bar{c}$)
\begin{equation}
I = \left|\frac{\partial\chi}{\partial c_m}\right|
\int {\cal D}\{c_m\} {\cal D}\{c\} {\cal D}\{\bar{c}\} {\cal D}\{a\}
\exp\left\{-\sum_{m\neq 0}c_m^2E_m - \pi c_0^2M -c\bar{c}M\right\}
\end{equation}
where we used $E_0=0$ and the fact that $\Psi_0$ is a normalized
eigenfunction, i.\ e.\ (from (\ref{2.3}) and (\ref{2.4}))
\begin{equation}
\Psi_0 = \frac{1}{\sqrt{M}}\frac{\partial\phi_c}{\partial\tau}, \quad
\sqrt{M}\Psi_0 = \frac{d\phi_c}{d a}
\end{equation}
where $M\equiv S_c=\frac{8}{g^2}$ is the classical action or instanton mass
(as it is also called). Integrating out all variables (and recalling that
${\cal D}\{a\} \rightarrow da$) the final result is
\begin{eqnarray}
I & = & 2T \left|\frac{\partial\chi}{\partial c_m}\right|
\prod_{m\neq 0}\sqrt{\frac{\pi}{E_m}}\frac{1}{\sqrt{M}}M \nonumber\\
& \equiv & 2TI_0 \sqrt{M}
\label{4.8}
\end{eqnarray}
where
\begin{equation}
I_0 = \left|\frac{\partial\chi}{\partial c_m}\right|
\prod_{m\neq 0}\left(\frac{\pi}{E_m}\right)^{\frac12}.
\label{4.9}
\end{equation}
In $I$ the factor $\frac{1}{\sqrt{M}}$ comes from the ${\cal D}\{c_0\}$
integration and $M$ from integrating out ${\cal D}\{c\} {\cal D}\{\bar{c}\}$.
We see therefore that the functional integration can be done without resorting
to the Faddeev--Popov method of inserting a delta function and interchanging
integration and limiting procedures.
The BRST procedure converts the ill defined integral
over the zero mode into a Gaussian integral and leads to the integration
over the instanton position, $da$, which gives rise to $2T$.

\section{The one--instanton transition amplitude and the contributions
from one instanton and an infinite number of instanton--antiinstanton pairs}

The following calculations are similar to those of the level splitting for the
double--well potential\cite{7} and of the decay rate for the inverted
double--well potential\cite{9}. The shift method which is effectively
defined by the
transformation
\begin{equation}
\chi(\tau) = \xi(\tau) \int_{-T}^T d\tau' \frac{\dot{y}(\tau')}{\xi(\tau')}
\label{5.1}
\end{equation}
where $\xi = \frac{d\phi_c}{d\tau}$
converts the functional integral into Gaussian form along with boundary
constraints which are then implemented with a delta function trick
analogout to the Faddeev--Popov method which then leads to a finite result
\cite{DittrichReuter}.
Taking care of the boundary condition
constraint, the functional integral for the field fluctuations is given by
\begin{equation}
I = \sqrt{\frac{1}{2\pi\xi(T)\xi(-T)
\int_{-T}^T  \frac{d\tau}{\xi^2(\tau)}}}.
\label{5.2}
\end{equation}
Considering the large time limit one finds
\begin{equation}
I \stackrel{T\rightarrow\infty}{\longrightarrow} \sqrt{\frac{1}{2\pi}}.
\label{5.3}
\end{equation}
Comparing with eq. (\ref{3.9}) it is seen that
\begin{equation}
I_0 =  \left|\frac{\partial\chi}{\partial c_m}\right| \prod_{m\neq 0}
\sqrt{\frac{\pi}{E_m}} = \frac{1}{\pi}\sqrt{\frac{E_0}{2}}.
\label{5.4}
\end{equation}
The eigenvalue $E_0$ vanishes only asymptotically when time $T$ tends to
infinity; its finite value corresponds to a finite time interval. $E_0$ can be
evaluated with a so--called boundary perturbation method\cite{7,9} which
gives rise to a formula for the ``unrenormalized'' eigenvalue--zero
eigenfunction $\xi = \frac{d\phi_c}{d\tau}$,
\begin{equation}
\xi(T)\frac{d\xi(T)}{d\tau} - \xi(-T)\frac{d\xi(-T)}{d\tau} =
E_0 \int_{-T}^T \xi^2 d\tau.
\label{5.5}
\end{equation}
Again we take the large time limit; the quasi zero eigenvalue due to the size
effect is obtained from eq.\ (\ref{5.5}) as
\begin{equation}
E_0 = 4e^{-2T}.
\label{5.6}
\end{equation}
Replacing $I_0$ in eq.\ (\ref{4.8}) by eq.\ (\ref{5.4}) with the value  $E_0$
of eq.\ (\ref{5.6}), the fluctuation part of the transition amplitude of the
one--instanton sector is seen to be
\begin{equation}
I^{(1)} = 2T\sqrt{M}\frac{\sqrt{2}}{\pi}e^{-T}.
\label{5.7}
\end{equation}

The contribution stemming from one instanton together with
an instanton--antiinstanton pair
can be calculated with the help of the group property for propagators
\cite{17}. The result is
\begin{equation}
I^{(3)} = \frac{(2T)^3}{3!}[\sqrt{M}]^3
\left[\frac{\sqrt{2}}{\pi}\right]^3\Delta^2 e^{-T}
\label{5.8}
\end{equation}
where the determinant
\begin{equation}
\Delta = \left[\frac{\pi}{\frac12 \frac{\partial^2 S_c}{\partial \phi_f^2}}
\right]^\frac12 \stackrel{T\rightarrow\infty}{\longrightarrow}\sqrt{2\pi}
\label{5.9}
\end{equation}
is determined from the end point integration of the group property of the
propagator and is evaluated with a formula given in the literature\cite{3} by
carefully taking the large time limit. The contribution from one instanton
plus $n$ pairs is a straightforward extension of eq.\ (\ref{5.8}) and is found
to be
\begin{equation}
I^{(2n+1)} = \frac{(2T)^{2n+1}}{(2n+1)!}(\sqrt{M})^{2n+1}
\left(\frac{\sqrt{2}}{\pi}\right)^{2n+1}\Delta^{2n}e^{-T}.
\label{5.10}
\end{equation}
The final result of the propagator is
\begin{equation}
\langle \phi_f,T | \phi_i,-T \rangle = \frac{1}{\sqrt{2\pi}}e^{-T}
\sinh\left[2T\frac{1}{\sqrt{2\pi}}\left(\frac{2^5}{g^2}\right)^{\frac12}
e^{-\frac{8}{g^2}}\right].
\label{5.11}
\end{equation}
Comparing with eq.\ (\ref{2.23}) we obtain the level shift, which is half of
the
width of the lowest energy band, i.\ e.\
\begin{equation}
\Delta {\cal E} = J =
\frac{1}{\sqrt{2\pi}}\left(\frac{2^5}{g^2}\right)^{\frac12}
e^{-\frac{8}{g^2}}
\label{5.12}
\end{equation}
which is in exact agreement with the result in the literature\cite{18}. The
periodic
potential has also been dealt with in the literature\cite{19} but the width of
the energy
band has not been given explicitly there.

\nonumsection{Acknowledgement}

This work was supported in part by the European Union under the Human
Capital and Mobility Programme. One of us (J.--G. Z.) thanks the
Alexander von Humboldt Foundation for support in the form of an Alexander
von Humboldt research fellowship.

\nonumsection{References}


\begin{thebibliography}{99}

\bibitem{1} A. Ringwald, Nucl.\ Phys.\ {\bf B330}(1989)1; O. Espinosa, ibid.\
{\bf B343}(1990)310.

\bibitem{2} S. Yu.\ Khlebnikov, V. A. Rubakov and P. G. Tinyakov,
Nucl.\ Phys.\ {\bf B367}(1991)334.

\bibitem{3} J.--Q. Liang and H. J. W. M\"uller--Kirsten, Phys.\ Rev.\
{\bf D46}(1992)4685; J.--Q. Liang and H. J. W. M\"uller--Kirsten, ibid.\
{\bf D50}(1994)6519; J.--Q. Liang and H. J. W. M\"uller--Kirsten, ibid.\
{\bf D51}(1995)718.

\bibitem{4} N. S. Manton and T. S. Samols, Phys.\ Lett. {\bf B207}(1988)179.

\bibitem{5} J.--Q. Liang, H. J. W. M\"uller--Kirsten and D. H. Tchrakian,
Phys.\
Lett.\ {\bf B282}(1992)105.

\bibitem{6} L. D. Faddeev and V. N. Popov, Phys.\ Lett.\ {\bf B25}(1967)29.

\bibitem{7} E. Gildener and A. Patrascioiu, Phys.\ Rev.\ {\bf D16}(1977)423.

\bibitem{8} S. K. Bose and H. J. W. M\"uller--Kirsten, Phys.\ Lett.\
{\bf A162}(1992)79.

\bibitem{9} J.--Q. Liang and H. J. W. M\"uller--Kirsten, Phys.\ Rev.\
{\bf D48}(1992)2963; {\bf 49}(1993)964(E).

\bibitem{10} J. Kurchan, D. R. Bes and S. Cruz Barrios, Phys.\ Rev.\
{\bf D38}(1988)3309.

\bibitem{11} J.\ Alfaro and P.\ H.\ Damgaard, Ann.\ Phys.\ {\bf 202}(1990)398.

\bibitem{12} H. J. W. M\"uller--Kirsten and Jian--zu Zhang, Phys.\ Rev.\
{\bf D50}(1994)6531; H. J. W. M\"uller--Kirsten and Jian--zu Zhang, Phys.\
Lett.\  {\bf B339}(1994)65.

\bibitem{13} V. A. Rubakov, D. T. Son and P. G. Tinyakov, Nucl.\ Phys.\
{\bf B404}(1993)65.

\bibitem{14} J.--Q. Liang and H. J. W. M\"uller--Kirsten, Phys.\ Lett.\
{\bf B332}(1994)129.

\bibitem{15} L. Baulieu and M. Bellon, Phys.\ Lett.\ {\bf B202}(1988)67.

\bibitem{16} J.--G. Zhou, F. Zimmerschied,
J.--Q. Liang, H. J. W. M\"uller--Kirsten
and D. H. Tchrakian, KL--TH--95/6.

\bibitem{17} R. H. Feynman and A. R. Hibbs, ``Quantum Mechanics and
Path integrals'' (McGraw--Hill, 1965).

\bibitem{DittrichReuter} W. Dittrich and M. Reuter, ``Classical and quantum
dynamics''
(Springer, 1992).

\bibitem{18} P. Achuthan, H. J. W. M\"uller--Kirsten and A. Wiedemann,
Fortschr.\ Phys.\ {\bf 38}(1990)77; R. B. Dingle and  H. J. W. M\"uller,
J. R. Angew.\ Math.\ {\bf 211}(1962)11; R. B. Dingle and  H. J. W. M\"uller,
J. R. Angew.\ Math.\ {\bf 216}(1964)123.

\bibitem{19} I. Bender, D. Gromes H. J. Rothe and K. D. Rothe, Nucl.\ Phys.\
{\bf B136}(1978)259.

\end{thebibliography}
\end{document}